# Comprehensive technology study of radiation hard LGADs

E – L Gkougkousis[a], L. Castillo Garcia[b], S. Grinstein[b], V. Coco[a]

[a] EP-LBD, CERN, Route du Meyrin 285, CH-1211 Genève 23, Switzerland
[b] IFAE, Edifici Cn, Campus UAB 08193 Bellaterra (Barcelona), Spain

Corresponding Author: egkougko@cern.ch

**Abstract**: Towards radiation tolerant sensors for pico-second timing, several dopants are explored. Using a common mask, CNM produced LGADs with boron, boron + carbon and gallium gain layers are studied, under neutron and proton irradiation. With fluences ranging from $10^{14}$ to $6 \times 10^{15}$ $n_{eq}/cm^2$ on both particle species, results focus on acceptor removal, gain reduction via electrical characterization, timing performance and charge collection. An emphasis is placed on stability via dark rate and operating voltage studies.

## 1. Introduction

Towards HL-LHC upgrades, both ATLAS and CMS envisage fast timing detectors for primary vertex separation and jet flavor tagging. Requirements vary from a per hit time resolution of *35 - 70 ps* with a radiation hardness of *2.5×$10^{15}$ $n_{eq}/cm^2$* at *2.4 < |η| < 4.0* for ATLAS [1], to a *30 ps* per track timing with a radiation tolerance of *1 × $10^{15}$ $n_{eq}/cm^2$* at *1.6 < |η| < 3.0* for CMS. Low Gain Avalanche Detectors (LGADs) [2] are the technology of choice for both experiments. In such a process, the addition of a secondary *p+* implant under the top *n+* electrode, referend to as multiplication layer, results in an abrupt variation of the electric field close to the anode [3]. This geometry induces charge multiplication though impact ionization at the multiplication layer.

## 2. Radiation effects on LGADs

Several radiation related mechanisms degrading LGAD performance can be identified. At the substrate level, trapping leads to reduced carrier lifetime compromising primary charge generation. Carrier lifetime *τ* is inversely proportional to radiation fluence (*Φ*) as described in (1).

$$1/\tau = \beta \times \Phi \qquad (1)$$

Additionally, deep acceptor level introduction at high fluences, modifies the effective doping ($N_{eff}$), leading to potential gain from the bulk, following (2).

$$N_{Eff.} = G \times \Phi \qquad (2)$$

Fluence dependence of both these effects can be described by the *β* and *G* coefficients as defined by the ROSE collaboration [4].

Multiplication layers are also affected by radiation. Active implant ($N_{G0}$) reduction with fluence is observed, described by the acceptor removal coefficient *c* as defined in (3) [5].

$$N_{G_\Phi} = N_{G_0} e^{-c\Phi} \qquad (3)$$

Finally, additional trapping and reduced mobility close to the gain layer further limit impact ionization. Through defect kinetics and interstitial diffusion, transformation of substitutional boron of the *p+* multiplication layer to interstitial and subsequent formation of *$B_iO_i$* complexes (4) has been proven to be detrimental due to the trapping properties of the latter [6]. Introduction of carbon can disrupt this mechanism by engaging oxygen in the formation of *$C_iO_i$* defects as detailed in (5). Replacement of boron with gallium, was also though to improve radiation hardness due to reduced gallium interstitial mobility, because of their higher mass and reduced defect formation cross-section.

$$Radiation + Si_s \rightarrow Si_i + B_s \rightarrow B_i + O \rightarrow B_iO_i \quad (4)$$

$$Radiation + Si_s \rightarrow Si_i + B_s \rightarrow C_i + O \rightarrow C_iO_i \quad (5)$$

In this study, an evaluation of the radiation hardness of boron (B), boron + carbon (B+C) and gallium (Ga) gain layer is presented, and the various radiation damage mechanisms are examined. Samples were fabricated on *4"* high resistivity (*~2 kΩ×cm*) Si-on-Si wafers of *50 μm* active thickness on a *300 μm* support wafer, following a "box-type" implant approximation, where the *p*-multiplication layer is adjacent to the *n* layer (figure 1). Single diodes with an active surface of *1 x 1 mm²* are used, surrounded by a deep *n+* junction terminator extension (JTE), a *p-stop* region and an *n+* guard ring at the front side. A uniform *p+* back-side contact is implemented (figure 2). Six fluence points are studied, from $1 \times 10^{14}$ to $6 \times 10^{15}$ $n_{eq}/cm^2$, for *24 GeV/c* CERN IRRAD protons and JSI fast neutrons. No thermal annealing was performed. All hereafter presented measurements were performed at three distinct temperature points: *-10°C*, *-20°C* and *-30°C*.

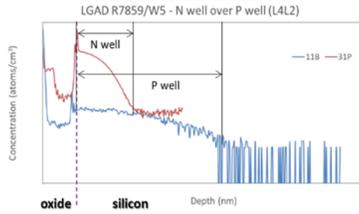
**Figure 1.** Secondary Ion Mass Spectr. doping profile of n⁺ *anode & p⁺* multiplication layer.

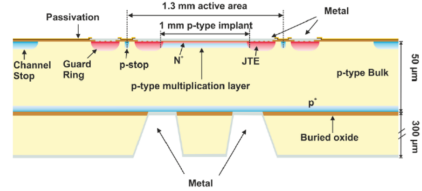
**Figure 2.** Cross-section of a Si-on Si wafer bonding LGAD diode geometry.

## 3. Active multiplication layer – The "Leakage Current Transition" method (LCT)

Evaluation of the radiation-induced multiplication layer degradation can be performed by studying the gain layer depletion voltage ($V_{GL}$), under a linear dependence assumption. While C-V studies for heavily irradiated sensors ($> 1 \times 10^{15}$ $n_{eq}/cm^2$) remain challenging [7], $V_{GL}$ can also be obtained through I-V analysis. In a "box-type" implant approximation, as previously defined, the Schottky-Mott equation predicts a flexion point of the I-V at the gain layer depletion voltage. The numerical derivative of the I-V at this point would pertain to a delta transition, convoluted with the doping profile mixing function (e.g., deviation form a pure box-model) and the measurement resolution. Consequently, $V_{GL}$ can be extracted by a gaussian fit at the I-V (figure 3). It should be noted that such an approximation - referred to as the "Leakage Current Transition method – LCT – is only valid for contiguous gain-n+ implants under a gaussian dopant-mixing assumption.

Applying LCT for different temperature points (*-10°C, -20°C, -30°C*), a $V_{GL}$ can be established for each sample and radiation fluence. Active multiplication implant is estimated (figure 4) by assuming a linear dependence between the reduction of $V_{GL}$ with fluence and remaining gain layer at each point. Acceptor removal coefficients (*c*) can be extracted for each radiation type and implant (Table 1) by fitting (3) to each one of the data series on figure 4. No significant variation of acceptor removal coefficient is observed between different multiplication layer implants. For all types, gain layer degradation follows an identical trend with respect to fluence both for neutron and non-NIEL normalized proton fluences. Ga samples present slight advantage under proton irradiation due to the more localized effect of induced defects. Higher mass associated with Ga, makes it less mobile in localized, collision type defects, accounting for marginal improvement under proton irradiation.

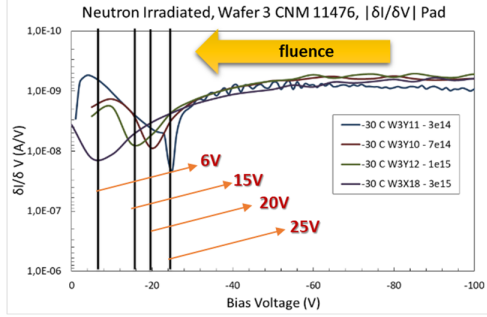

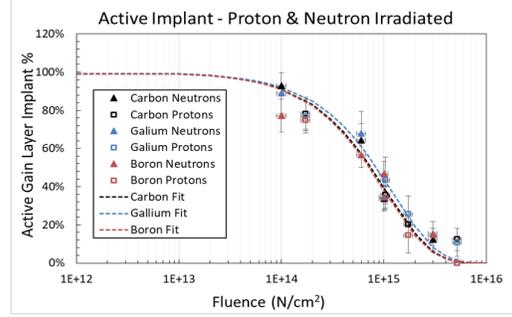

**Figure 3.** Numerical derivative of pad I-V for different fluences with visible the position of the gain layer depletion voltage.

**Figure 4.** Relative active implant for all gain layer configurations at different fluences. No NIEL normalization applied.

**Table 1.** Acceptor removal coefficients for $n^0$ and $p^+$ irradiated samples on all gain layers

| Gain layer | Effective gain removal coefficient | | |
| --- | --- | --- | --- |
| implementation | $n^0$ irradiated | $p^+$ irradiated | Combined |
| **Gallium** | $(8.3 \pm 1.2) \times 10^{-16}$ | $(1.4 \pm 0.19) \times 10^{-15}$ | $(8.2 \pm 0.8) \times 10^{-16}$ |
| **Boron + Carbon** | $(8.8 \pm 0.9) \times 10^{-16}$ | $(1.70 \pm 0.22) \times 10^{-15}$ | $(9.3 \pm 0.8) \times 10^{-16}$ |
| **Boron** | $(8.2 \pm 1.3) \times 10^{-16}$ | $(1.96 \pm 0.16) \times 10^{-15}$ | $(9.69 \pm 0.10) \times 10^{-16}$ |

## 4. Effective Gain – GR vs Pad Method (GPM)

Through the LCT approach, gain layer de-activation via acceptor removal may be probed but, trapping effects and bulk acceptor level introduction, are ignored. The effect of all three mechanisms is expressed by estimating the effective gain, thought comparison of a non-multiplication vs a multiplication region at the same die. Because of the unfirmly implanted back side and the p-stop GR – pad isolation, two separate diodes can be considered at the same die. The first, formed between the pad active area and backside, includes the multiplication layer. The second, defined between the GR n-well and p-backside implant, is a PIN diode. Leakage currents of the two regions can be expressed as:

$$I_{pad}^{\Phi=0} = S \times I_s \times \left(e^{\frac{eV}{nkT}} - 1\right) \times G(e^V, T, 0) \quad and \quad I_{GR}^{\Phi=0} = I_s \times \left(e^{\frac{eV}{nkT}} - 1\right) \quad (6)$$

where $S$ represents a geometry factor accounting for the difference in volume and shape of the two depleted regions, $I_s$ is the generation current and $G(e^V,T,\Phi)$ the gain dependent contribution.

Following the alpha factor formalism, the radiation indued increase of the generation current portion can be expressed as $\Delta I_s = \alpha \times \Delta \Phi$. By dividing the leakage currents of the two regions and normalizing at $\Phi = 0$, a gain-only dependent ratio can be obtained (7):

$$\left|I_{pad}^{\Phi}/I_{GR}^{\Phi}\right| = S \times G(e^V, T, \Phi) \qquad (7)$$

The $I_{pad}/I_{GR}$ ratio can be arithmetically calculated (figure 5). The exponential behavior with respect to bias voltage follows a $I_{pad}/I_{GR} = m \times b^V$ formalism. In such an approach, $b$ represents the gain induced increase of $I_{pad}/I_{GR}$ and can be extrapolated by a linear fit at the semi-logarithmic plane.

The temperature dependence of gain imposes an equivalent dependence of $b$. Normalization with respect to the unirradiated sample is therefore applied separately for each temperature. The effective gain for all fluences and doping configurations is estimated individually for proton and neutron irradiation (figures 6 & 7). This approach will be thereafter referred to as the "Guard Ring vs Pad Method" – GPM.

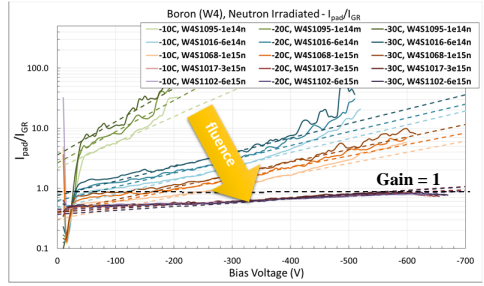

**Figure 5.** $I_{pad}/I_{GR}$ versus bias voltage at the semi-log plane with the corresponding exponential fits for different temperatures and fluences. A slope (*b* factor) reduction is observed with increased fluence.

**Table 2.** Effective gain removal coefficient for $n^0$ and $p^+$ irradiated samples on all gain layers

| Gain layer implementation | Effective gain removal coefficient | |
|---|---|---|
|  | $n^0$ irradiated | $p^+$ irradiated |
| **Gallium** | $(3.01 \pm 0.9) \times 10^{-14}$ | $(2.02 \pm 0.11) \times 10^{-14}$ |
| **Boron + Carbon** | $(2.57 \pm 1.1) \times 10^{-15}$ | $(1.37 \pm 0.24) \times 10^{-14}$ |
| **Boron** | $(2.25 \pm 0.39) \times 10^{-14}$ | $(2.25 \pm 0.28) \times 10^{-14}$ |

Using the modified effective doping formulation as defined by (8) the removal coefficients *c'* can be estimated (table 2) for all gain layer implementations. Identical acceptor re-introduction rates $g_c$ and initial concentrations $N_{eff,0}$ are assumed for all gain layer variations as structures are implemented in identical substrates. As reported in table 2, carbonated wafers present at least two times lower removal coefficients under proton irradiation with respect to boron/gallium and up to 7 times under neutron. It is noticeable that estimated effective gain removal coefficients are up to 10x higher than previously calculated acceptor removal coefficients.

$$N_{eff}(\Phi) = N_{eff_0} - N_C(1 - e^{-c'\Phi}) + g_c\Phi \qquad (8)$$

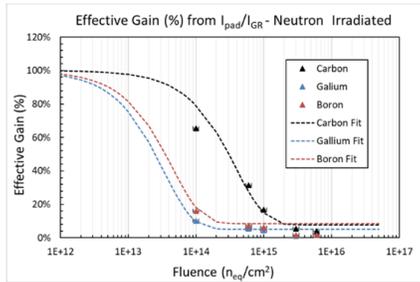
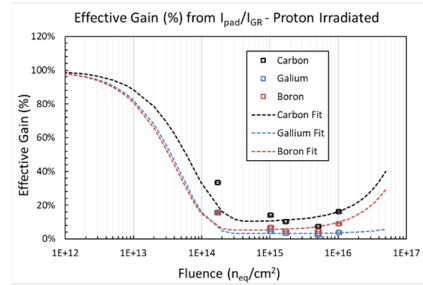

**Figure 6.** Effective gain vs fluence for $n^0$ irradiated B, B+C & Ga gain layers.

**Figure 7.** Effective gain for $p^+$ irradiated samples. The increase at high fluences is a result of deep acceptor level re-introduction.

## 5. Real Gain estimation – Charged particle and timing measurements

For a complete evaluation or all radiation damage effects, Minimum Ionizing Particles (MIPs) induced measurements are required. β-particles from a *$^{90}$Sr* source are used and charge, time resolution, noise and rise time are evaluated at different bias voltages for all samples. Two-object coincidence triggering is implemented, using a known time-resolution LGAD as a reference. Care is taken to exclude the *$^{90}$Y* lower energy contribution using a *90 μm* aluminum collimator while a high statistic of 5000 events per point is recorded. Collected charge is defined as the MPV of the Landau X Gauss convoluted fit of the charge distribution of all recorded events per point. A time-walk correction via constant fraction discriminator is also utilized through CFD percentage optimization (figure 8).

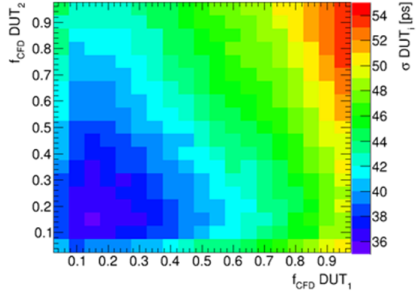

**Figure 8.** 2D - CFD optimization map with time resolution at the colour axis for two identical unirradiated LGADs.

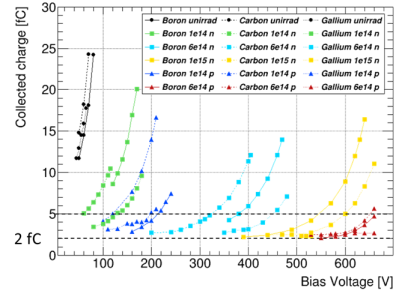

**Figure 9.** Collected charge for $n^0$ and $p^+$ irradiated samples for all gain layer configurations at -30 ºC.

An identical behaviour is observed prior to irradiation for all three gain layer implementations across all temperatures (figure 9). On neutron irradiated samples, a *15%* increase on the bias is required for gallium implanted samples to achieve equivalent collected charge with respect to boron ones. The trend persists across all fluences, irradiation species and temperatures. For carbonated samples, an opposite behaviour is observed. Same charge levels are obtained at a *20%* lower biasing with respect to boron samples, across all fluences, irradiation types and temperatures. For all doping types, $p^+$ irradiated samples require a factor of two increase in bias votlage to reach equivalent perfromance with respect to $n^0$ irradiation. The sae picture, in regards of doping type, temperature and irradiation type are also reflected on time resolution. It must be noted that all samples achieve the *30 - 70 ps* requirements imposed by both ATLAS and CMS experiments (figure 10). Further signal analysis does not reveal any pulse-shape variations realted to doping type as can be seen for example in the rise time (figue 11).

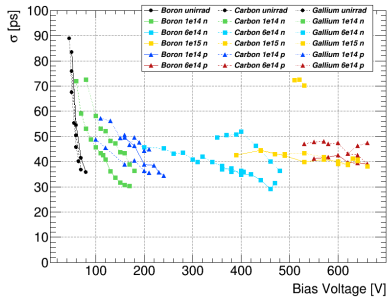

**Figure 10.** Timing resolution of all gain configuration at difference fluences.

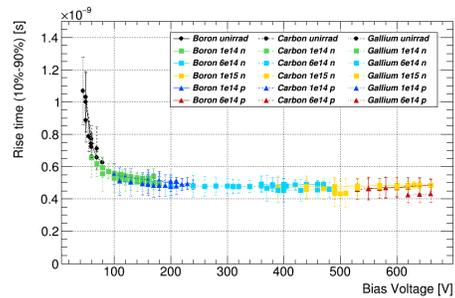

**Figure 11.** Rise time for different gain layer configurations and irradiation level.

## 6. Comparative Studies
Evaluation of possible effects in leakage current across different gain layer types is caried out though I-V analysis. A comparison of non-gain regions between carbonated and standard boron samples reveals a *38%* increase of leakage current in the carbon case. Such an increase is unrelated to the enhanced gain observed for carbonated devices after irradiation and is consistently present for fluences of up to $1 \times 10^{15}$ $n_{eq}/cm^2$ on both $p^+$ and $n^0$ irradiated samples across all temperatures (figure 12). At higher fluences, increase due to radiation induced damage current prevails, rendering prohibitive any meaningful comparison.

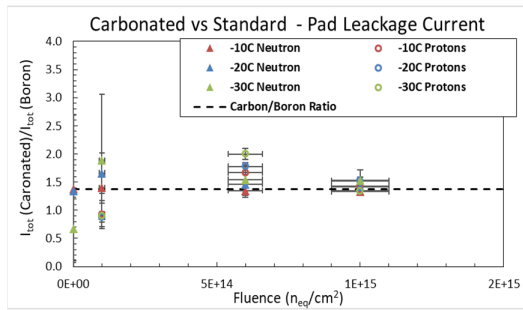 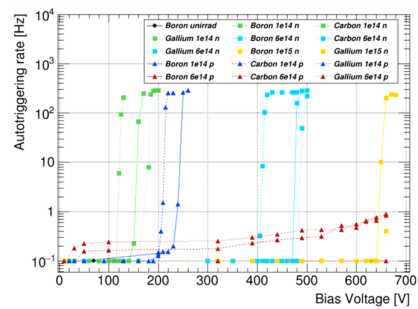

**Figure 12.** Leakage current ratio of carbonated vs non-carbonated samples.

**Figure 13.** S-stability curves for all samples at $-30^oC$.

Stability of all devices is evaluated though a dark rate study. With no external stimulus, four continuous self-trigger events within a time frame of up to *5 min* are recorded and a median rate is extracted. The process is repeated 1000 times for each voltage point. A distribution of median rates is obtained and a poissonian fit is applied. The extrapolated from fit median rate is then regarded as the dark rate value corresponding to the specified point. Through bias voltage scanning, s-like stability curves are obtained for each sample (figure 13). An exponential increase in dark rate is observed at bias voltages up to *10-15 V* lower than the established breakdown point of the device. For B+C samples, a *20%* decrease of the stable operating voltage point is noted with respect to B multiplication layer. On the contrary, for gallium implanted samples, the situation is inversed with a *10%* increase at the operation point. Such behavior is in agreement with the charge collected measurements.

## 7. Conclusions

A systematic study of B, C+B and Ga multiplication layers was presented, and different radiation damage mechanisms probed. In terms of active gain layer, no variations were observed within uncertainty limits. Effective gain of B+C devices was proven to be at least 2 times higher with respect to B samples while, Ga devices were found to be marginally worse. Evaluating charge collection and timing, a *20%* improvement is observed by C introduction and a *15%* degradation is noted for Ga. As adverse effects, a *33%* increase in leakage current, a *20%* lower stability point and headroom allowing *100%* efficient operation up to $2 \times 10^{15}$ $n_{eq}/cm^2$ are observed for carbonated devices.